\documentclass[aps, prl, superscriptaddress, reprint, amsmath,amssymb]{revtex4-1}
\usepackage{graphicx}
\usepackage{dcolumn}
\usepackage{bm}
\newcommand{\Spi}{Sierpi\'{n}ski }

\begin{document}

\title{$p$-band engineering in artificial electronic lattices}

\author{M. R. Slot}
\thanks{Both authors contributed equally.}
\affiliation{Debye Institute for Nanomaterials Science, Utrecht University, Utrecht, Netherlands}
\author{S. N. Kempkes}
\thanks{Both authors contributed equally.}
\affiliation{Institute for Theoretical Physics, Utrecht University, Utrecht, Netherlands}
\author{E. J. Knol}
\affiliation{Institute for Molecules and Materials, Radboud University, Nijmegen, Netherlands}
\author{W. M. J. van Weerdenburg}
\affiliation{Institute for Molecules and Materials, Radboud University, Nijmegen, Netherlands}
\author{J. J. van den Broeke}
\affiliation{Institute for Theoretical Physics, Utrecht University, Utrecht, Netherlands}
\author{D. Wegner}
\affiliation{Institute for Molecules and Materials, Radboud University, Nijmegen, Netherlands}
\author{D. Vanmaekelbergh}
\affiliation{Debye Institute for Nanomaterials Science, Utrecht University, Utrecht, Netherlands}
\author{A. A. Khajetoorians}
\affiliation{Institute for Molecules and Materials, Radboud University, Nijmegen, Netherlands}
\author{C. Morais Smith}
\affiliation{Institute for Theoretical Physics, Utrecht University, Utrecht, Netherlands}
\author{I. Swart}
\email{Correspondence to: I.Swart@uu.nl, C.deMoraisSmith@uu.nl}
\affiliation{Debye Institute for Nanomaterials Science, Utrecht University, Utrecht, Netherlands}

\date{\today}
\begin{abstract}
Artificial electronic lattices, created atom by atom in a scanning tunneling microscope, have emerged as a highly tunable platform to realize and characterize the lowest-energy bands of novel lattice geometries. Here, we show that artificial electronic lattices can be tailored to exhibit higher-energy bands. We study $p$-like bands in four-fold and three-fold rotationally symmetric lattices. In addition, we show how an anisotropic design can be used to lift the degeneracy between $p_x$- and $p_y$-like bands. The experimental measurements are corroborated by muffin-tin and tight-binding calculations. The approach to engineer higher-energy electronic bands in artificial quantum systems introduced here enables the realization of complex band structures from the bottom up.
\end{abstract}

\maketitle
Bands composed of orbitals beyond $s$-type play a key role for the electronic and magnetic properties of materials. Orbitals are characterized by the shape of the wave function, establishing a degree of freedom for the electrons in addition to spin and charge \cite{Tokura}. Higher orbitals (\textit{i.e.} beyond $s$-type) can give rise to interesting band structures, such as a Dirac cone and flat band for the $p$-orbital bands of a honeycomb lattice~\cite{Wu2007, Wu2008, Beugeling}. In addition, colossal magnetoresistance of several transition metal-oxides is related to the $d$-orbital bands in these systems~\cite{Tokura}. In the presence of interactions, the nodal character of bosons in a $p$-band condensate~\cite{Wirth, Oelschlaeger, Mueller, SmithHemmerich} or of the superconducting order parameter in unconventional superconductors~\cite{Kirtley, Maeno, Mackenzie, Kvorning} leads to interesting broken-symmetry quantum phases and novel quantum effects. 

Generally, the term 'orbitals' refers to the nodal structure of the wave functions at the lowest ($s$) and higher ($p$, $d$, etc.) energies, which is equivalent to the nodal structure of atomic orbitals. More specifically, $s$-orbitals in a lattice yield an either positive or negative wave-function amplitude at the lattice sites. In contrast, $p$-orbitals alternate from a positive contribution between lattice sites via a node at the site itself to a negative contribution on the other side. Bands originating from these $p$-orbitals, so-called $p$-like bands, were studied in optical \cite{Kock, Li, Lewenstein, Wirth, Oelschlaeger, Mueller, SmithHemmerich}, photonic \cite{Milicevic, Cantillano}, and polariton lattices~\cite{Jacqmin,Klembt,Whittaker}.

The manipulation capability of the low-temperature scanning tunneling microscope has been used to create atomically precise structures. This allows one to precisely control the electronic and spin coupling between atoms ~\cite{Nilius,Nilius2,Folsch,Hirjibehedin2006, Khajetoorians2012, Kamlapure}. For artificial electronic systems, periodic (e.g. honeycomb, Lieb, checkerboard~\cite{Manoharan, Drost2017, Slot2017, Otte}) and non-periodic (quasi-crystalline Penrose tiling and \Spi fractals~\cite{Collins2017, Kempkes}) geometries, as well as topologically non-trivial 1D chains ~\cite{Drost2017, Huda} have been made. However, all of the experiments on electronic lattices focused on the lowest energy bands, derived from $s$-like orbitals at the artificial-atom sites. In Ref. \cite{Ma}, Ma \textit{et al.} pointed out that the higher-energy bands found for the Lieb lattice in Ref. \cite{Slot2017} are well-described by $p$-like bands, supported by a tight-binding model and plane-wave calculations.

Here, we first experimentally identify $p$-like bands in a four-fold rotationally symmetric Lieb lattice in which all sites host degenerate $p_x$- and $p_y$-like orbitals, by comparing the observed spatially dependent local density of states (LDOS) with muffin-tin calculations. Because the artificial atoms are two-dimensional, there are only 2 $p$-like orbitals centered on each site, instead of 3 for real atoms. Then, we engineer a Lieb lattice in which the $p$-degeneracy at the edge sites is broken, resulting in a separate $p_x$-like ($p_y$-like) orbital at the $x$ ($y$) edge site in the energy range of interest. Next, by introducing an asymmetry in the lattice, we are able to lift the energy degeneracy of the remaining $p_x$- and $p_y$-like orbitals and independently access these different degrees of freedom upon tuning the energy. Finally, to illustrate that the $p$-like band description is also applicable to systems with other symmetries, we investigate $p$-like bands in the three-fold rotationally symmetric honeycomb lattice.

The electronic lattices are realized and characterized in two low-temperature scanning tunneling microscopes (STMs), located in Utrecht and Nijmegen (both Scienta Omicron LT-STM, $T\,=\,4.5\,$K), in ultrahigh vacuum ($p\approx 10^{-10}\,$mbar). Carbon monoxide (CO) molecules are leaked into the chamber and adsorbed onto a cold Cu(111) single crystal in the STM, cleaned by sputtering and annealing cycles, such that the surface coverage was approximately $0.3$ molecules per nm$^{2}$. A Cu-coated tungsten or platinum-iridium tip, prepared by gentle contact with the Cu(111) surface, is used for both the assembly and the characterization of the lattices. The CO molecules act as repulsive scatterers for the surface-state electrons of Cu(111) \cite{Paavilainen}. By positioning the CO molecules with atomic precision using the STM tip~\cite{Stroscio1991, Bartels, Celotta}, the electrons at the surface of the Cu(111) crystal are confined to the regions in between the molecules, leading to the formation of an electronic lattice. In this manner, the confined 2D electron gas (2DEG) at the surface is patterned to form "artificial atoms" with a specific size \textemdash and thus on-site energy \textemdash and a tunable coupling between them. We characterize the nodal plane character of the wave function by mapping the LDOS at constant height at various energies. We utilize a lock-in technique and apply a modulation to the sample bias (modulation amplitude 5-20 mV r.m.s., frequencies $273\,$Hz and $4321\,$Hz). As evidenced by previous work~\cite{Slot2017,Kempkes}, the muffin-tin model without many-body interactions yields a good description of the electronic band structure of artificial lattices generated by CO molecules on a Cu(111) surface. The muffin-tin calculations model the CO molecules by a circular repulsive potential with a radius $r=0.3\,$nm and height of $V_{\mathrm{CO}}=0.9\,$eV in the Schr\"odinger equation for the 2DEG~\cite{Park,Slot2017}. Not only the LDOS $|\Psi|^2$ is calculated, but also the sign of the contributing wave functions $\Psi$ is extracted. The results are further corroborated by single-particle tight-binding calculations, which model the artificial-lattice geometry with the appropriate orbitals at the artificial-atom sites (see Supplemental Material (SM)~\cite{Supp}).\\
\begin{figure}[t!]
\centering
\includegraphics[width=0.5\textwidth]{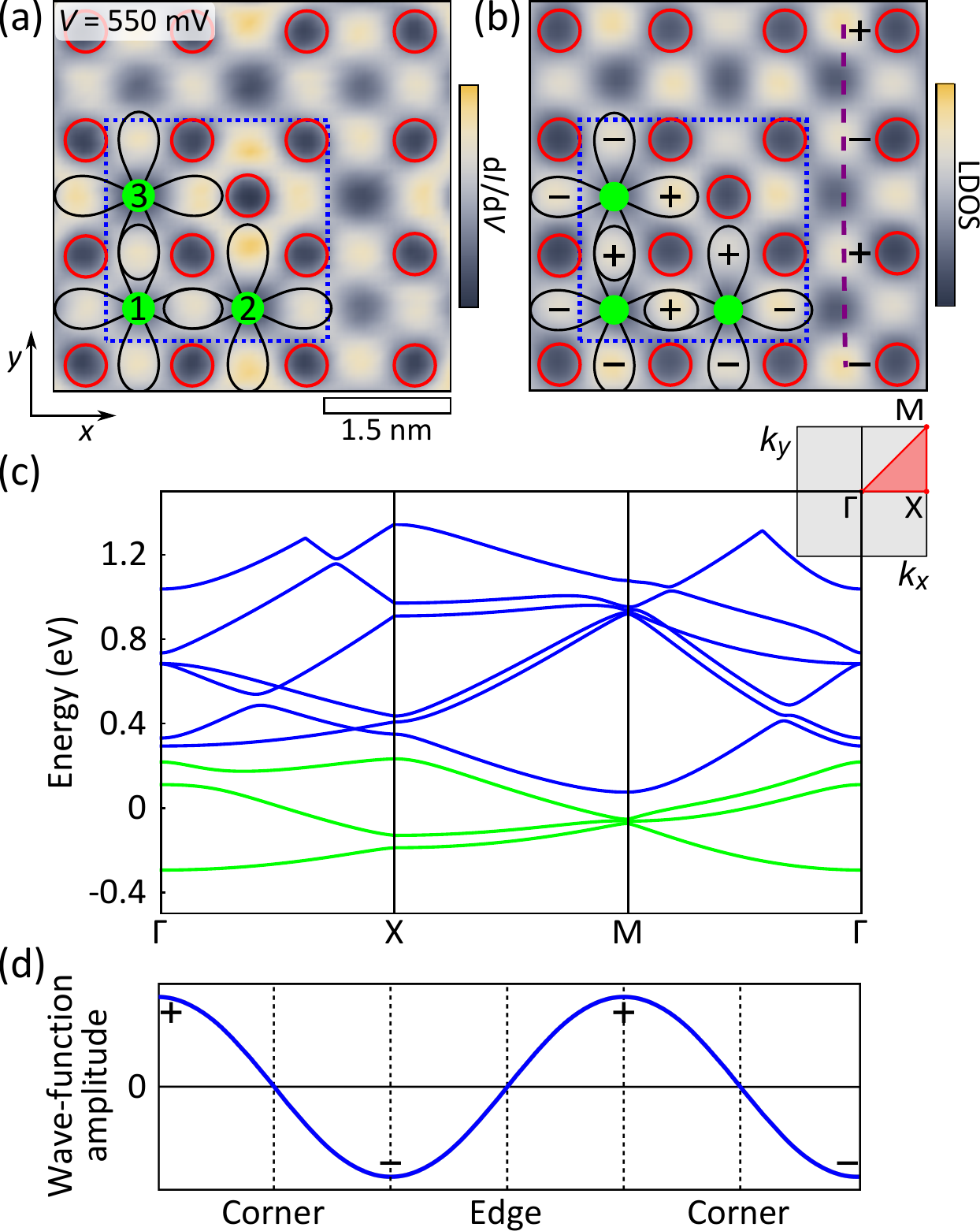}
\caption{(a) Differential-conductance map acquired at $V = 550\,$mV above a Lieb lattice, shown for part of the lattice. Images of the entire lattice can be found in the SM~\cite{Supp}. The unit cell (blue dashed box) contains 3 artificial-atom sites (green): one corner site (1) and two edge sites (2 and 3). The sites are generated by the configuration of CO molecules on a Cu(111) surface indicated by the red contours. The $p-$like orbitals in the unit cell are outlined by a black contour. The map was acquired at $+110\,$pm with respect to the setpoint $I_{\mathrm{set}} = 1\,$nA, $V_{\mathrm{set}} = 50\,$mV, with a bias modulation $V_{\mathrm{rms}} = 20\,$mV. 
(b) LDOS map for the same energy simulated using the muffin-tin model. The $p-$like orbitals and the sign of the wave function are indicated in black. (c) Band structure along the high symmetry directions of the Brillouin zone (see inset) as calculated by a muffin-tin approach. The 3 lowest $s-$like bands are indicated in green, the 6 p-orbital bands in blue. (d) Bloch-type wave function along the purple dashed line in the muffin-tin simulation in (b). The wave function exhibits a positive maximum or negative minimum between the sites, 
while it is zero at the artificial-atom sites (corner and edges).
}
\label{FIG1}
\end{figure}
First, we describe higher-orbital bands in a lattice with a four-fold rotational symmetry. For this purpose, we chose the Lieb lattice, which is a square-depleted lattice consisting of three (artificial) atoms (green) in a unit cell (blue dashed box) \cite{Weeks}, as illustrated in Fig.~1a. With three sites per unit cell this lattice provides more flexibility than a square lattice. The three inequivalent artificial-atom sites consist of one corner site (1) and two edge sites (2 and 3). The size of the unit cell is chosen to be $6\sqrt{3} a \times 10 a \approx 2.66\,$nm$\times2.56\,$nm, where $a = 0.256\,$nm is the Cu(111) nearest-neighbor distance (see SM~\cite{Supp}). This electronic lattice is realized by an array of CO molecules (red circles) on Cu(111), which acts as a repulsive potential and confines the electrons to the sites of the Lieb lattice \cite{Slot2017, Qiu}. At low energies $E < 200\,$meV, there are two bands exhibiting a Dirac cone at the corners of the Brillouin zone intersected by a third band (\textit{c.f.} green curves in Fig. 1c). These bands have been observed experimentally and are well-described by a tight-binding model using $s$-orbitals at the artificial-atom sites~\cite{Drost2017,Slot2017}. 

Fig.~1a shows a differential-conductance map acquired at a higher energy, $V = 550\,$mV. The map shows the central plaquette of a Lieb lattice of $5 \times 5$ unit cells (see SM~\cite{Supp}). We observe that the artificial-atom sites (green) exhibit nodes in the LDOS, while there is an enhanced LDOS between the sites. This is in agreement with the result from muffin-tin calculations (see Fig. 1b). The nodal pattern corresponds to that of $p$-like orbitals, indicated schematically by black contours. These two-dimensional $p$-orbitals consist of degenerate $p_x$- and $p_y$-like states, which extend between the sites to overlap with $p_{x,y}$-states localized on a neighboring site. The variations in LDOS maxima between dangling and overlapping $p$-like orbitals can be attributed to differences in confinement (see SM~\cite{Supp}).
We calculate the band structure of the Lieb lattice in the energy range $-0.5\,\mathrm{eV} < E < 1.4\,$eV using the muffin-tin model, as shown in Fig.~1c. For $-300\,\mathrm{meV} < E < 200\,$meV, we obtain the three previously described lowest-energy bands (green). At higher energies, $E > 200\,$meV, additional bands are predicted (blue). To study the nature of the bands in more detail, we extract the wave-function amplitudes from the muffin-tin calculations. Fig.~1d shows the value of the wave function along the purple line indicated in Fig.~1b. Note that the amplitude of the wave function changes sign at the positions of the artificial atoms. In between these sites, it is either positive or negative. This corresponds to overlap of lobes with the same sign, \textit{i.e.} a bonding combination of $p$-like orbitals on adjacent sites, as indicated in Fig. 1b. Since there are two artificial atoms per unit cell along the $x$ and $y$ direction, the periodicity of the wave in Fig. 1d is the same as that of the unit cell. Thus, the muffin-tin calculations confirm the typical $p$-like character of the wave functions. A similar analysis of the lower-energy bands shows that the associated wave functions have $s$-like character (see SM~\cite{Supp}). If the energy is increased further, anti-bonding $p_\pi$-like and higher-orbital bands occur (see calculations in SM~\cite{Supp}).

\begin{figure}[b!]
\centering
\includegraphics[width=0.5\textwidth]{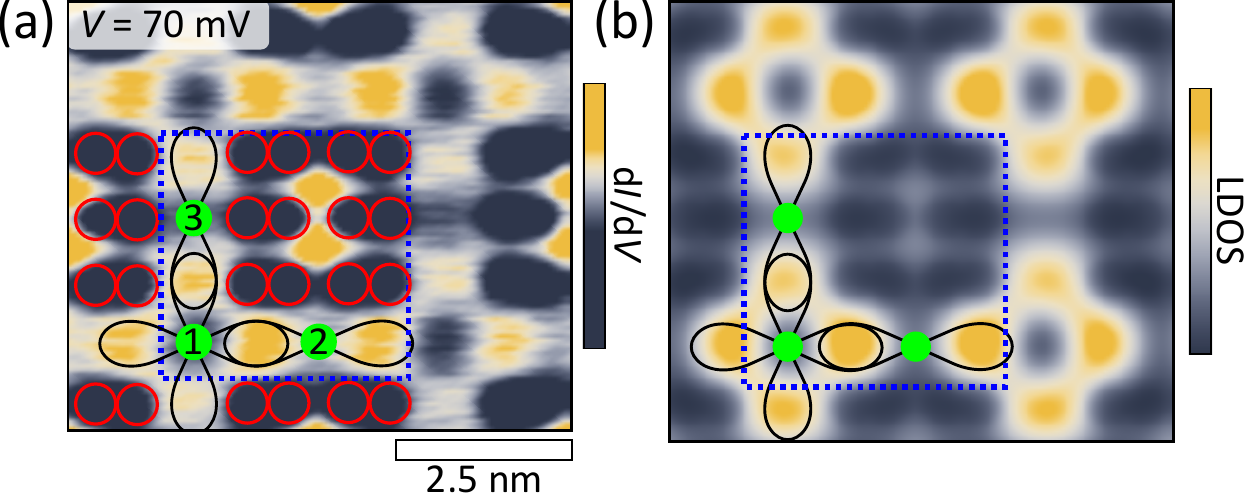}
\caption{(a) Differential-conductance map above a symmetric Lieb-like lattice acquired at $V = 70\,$mV, shown for part of the lattice (see SM~\cite{Supp} for entire lattice). The CO molecules are encircled in red and $p_x$- and $p_y$-like orbitals (black) are indicated in one unit cell (blue dashed box). The map was acquired after stabilization at $I_{\mathrm{set}} = 1\,$nA, $V_{\mathrm{set}} = 70\,$mV, with a bias modulation $V_{\mathrm{rms}} = 10\,$mV.
(b) LDOS map for the same energy simulated using the muffin-tin model.}
\label{FIG2}
\end{figure}

Next, we alter the design of the Lieb lattice for two purposes. First, we shift the $p$-bands down towards the Fermi energy, where the surface state of Cu(111) has a free-electron-like dispersion. Close to the Fermi energy the COs are more effective at confining the surface-state electrons and the influence of the bulk is minimized. We achieve this by increasing the size of the artificial-atom sites \cite{Manoharan}, reducing the confinement of the electrons. Second, we lift the degeneracy of the $p_x$- and $p_y$-like orbitals at the same edge sites by adding additional CO molecules, such that the dangling orbital is pushed to much higher energy. In Fig.~2a, we show a differential-conductance map of this modified Lieb-like lattice with larger unit cells. The unit cell with size $14 a \times 8 \sqrt{3} a \approx 3.58\,$nm$\times \,3.55\,$nm is indicated by a dashed blue box. It contains 12 CO molecules (encircled in red). Note that compared to the lattice discussed above, the edge sites in this design are more confined in the direction perpendicular to the line connecting two corner sites. As a guide to the eye, the locations of the artificial atoms are indicated in green and the $p$-like orbitals are outlined in black. For $-300\,\mathrm{meV} < E < -100\,$meV, the $s$-like bands were reproduced (see SM~\cite{Supp}). The differential-conductance map shown in Fig.~2a was acquired at $V = 70\,$mV. Note that there are nodes at the positions of the artificial atoms. The corner sites (1) exhibit both $p_x$- and $p_y$-orbitals. However, in contrast to the lattice in Fig.~1, the edge sites (2 and 3) exhibit only the $p_x$- (2) \emph{or} the $p_y$-like (3) state in the bonding direction at this energy. Indeed, the additional CO molecules in this design shift the dangling orbital perpendicular to this direction ($p_y$-like for site (2) and  $p_x$-like for site (3)) to much higher energy. Fig.~2b presents the corresponding LDOS calculated using the muffin-tin model. The nodal character is in excellent agreement with the experimental data. From the muffin-tin results, the wave-function amplitudes can again be extracted, corroborating the $p$-like character of the wave-functions (see SM~\cite{Supp}). Thus, by tailoring the design, it is possible to have only $p_x$-like orbitals on one site and $p_y$-like orbitals on another at a given energy.

\begin{figure}[t!]
\centering
\includegraphics[width=0.5\textwidth]{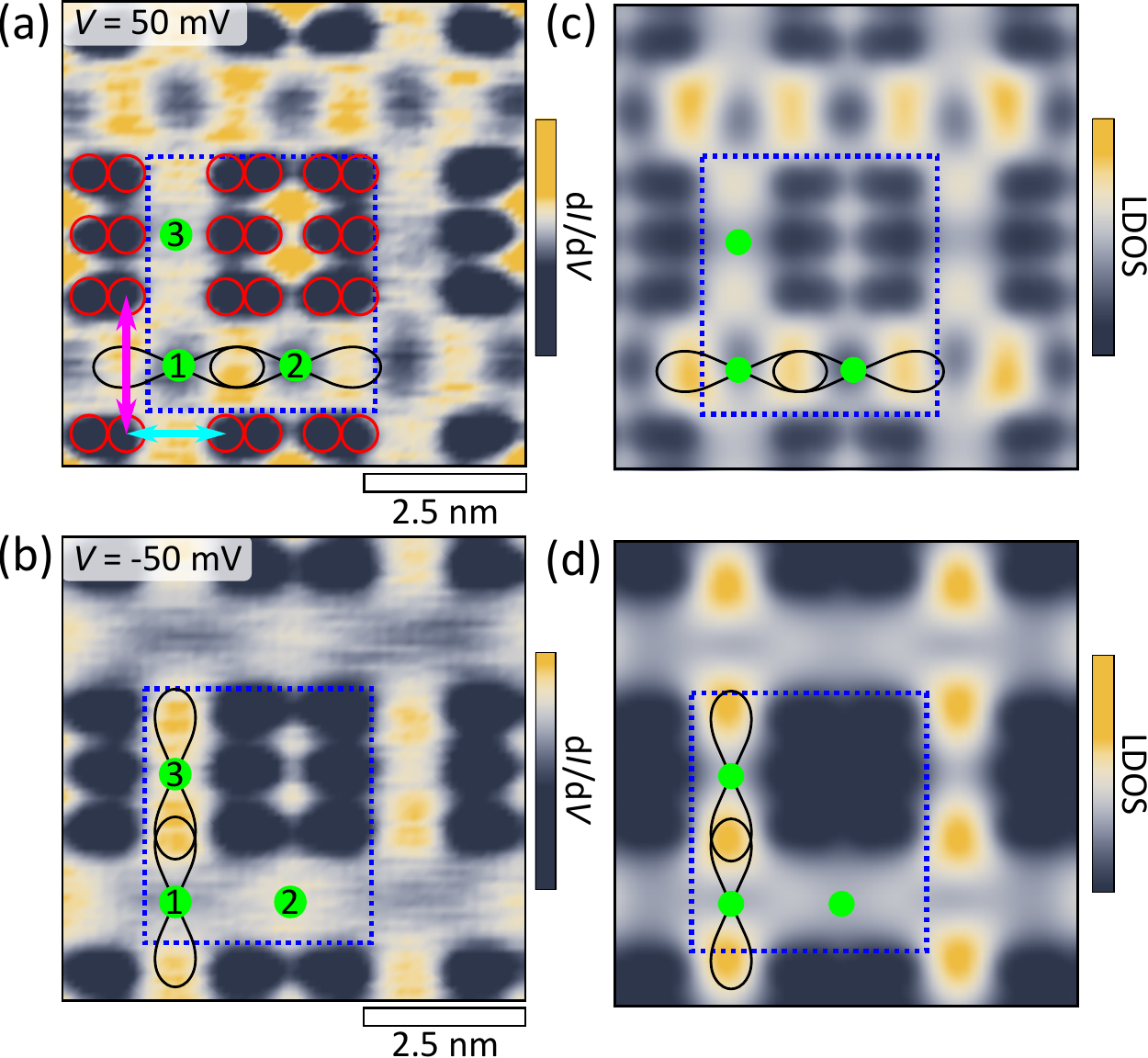}
\caption{(a-b) Differential-conductance maps acquired at $V = +50\,$mV and $V = -50\,$mV, respectively, above an asymmetric Lieb-like lattice with more spacing in the $y$-direction (pink arrow) than in the $x$-direction (cyan arrow). A part of the lattice is shown. The maps were acquired after stabilization at $I_{\mathrm{set}} = 1\,$nA, $V_{\mathrm{set}} = +50\,$mV and $-50\,$mV, respectively, with a bias modulation $V_{\mathrm{rms}} = 5\,$mV. 
(c-d) Same as (a-b), calculated using the muffin-tin model.}
\label{FIG3}
\end{figure}

We now turn our attention to lifting the energy degeneracy of the $p_x$- and $p_y$-like orbitals at all sites by introducing an asymmetry in the lattice sites~\cite{Menezes}. Specifically, we break the four-fold rotational symmetry by increasing the width of artificial-atom sites 1 and 2 in the $y$-direction by $\sqrt{3}a$ (pink arrow in Fig.~3a) while the $x$-direction remains the same (cyan arrow). This results in a unit-cell size of $14 a \times 9 \sqrt{3} a \approx 3.58\,$nm$\times 3.99\,$nm.
Figures~3a-b show differential-conductance maps of the asymmetric lattice at $+50\,$mV and $-50\,$mV, respectively. At +$50\,$mV, we only observe the $p_x$-like orbitals (Fig.~3a). In contrast, at $-50 \,$mV, only $p_y$-like orbitals contribute to the image contrast (Fig.~3b). This can be ascribed to the larger amount of space and therefore reduced confinement in the $y$-direction. The energy splitting can be considered as an artificial-lattice analogue of crystal-field splitting in solid-state materials. The measurements are reproduced by muffin-tin calculations (see Figs. 3c and 3d, and the SM~\cite{Supp}). Note that at $-50\,$mV, we observe a contribution of the $s$-orbitals at sites 2 and 3 in addition to the $p_y$-orbitals at site 3. 
\begin{figure}[t!]
\centering
\includegraphics[width=0.5\textwidth]{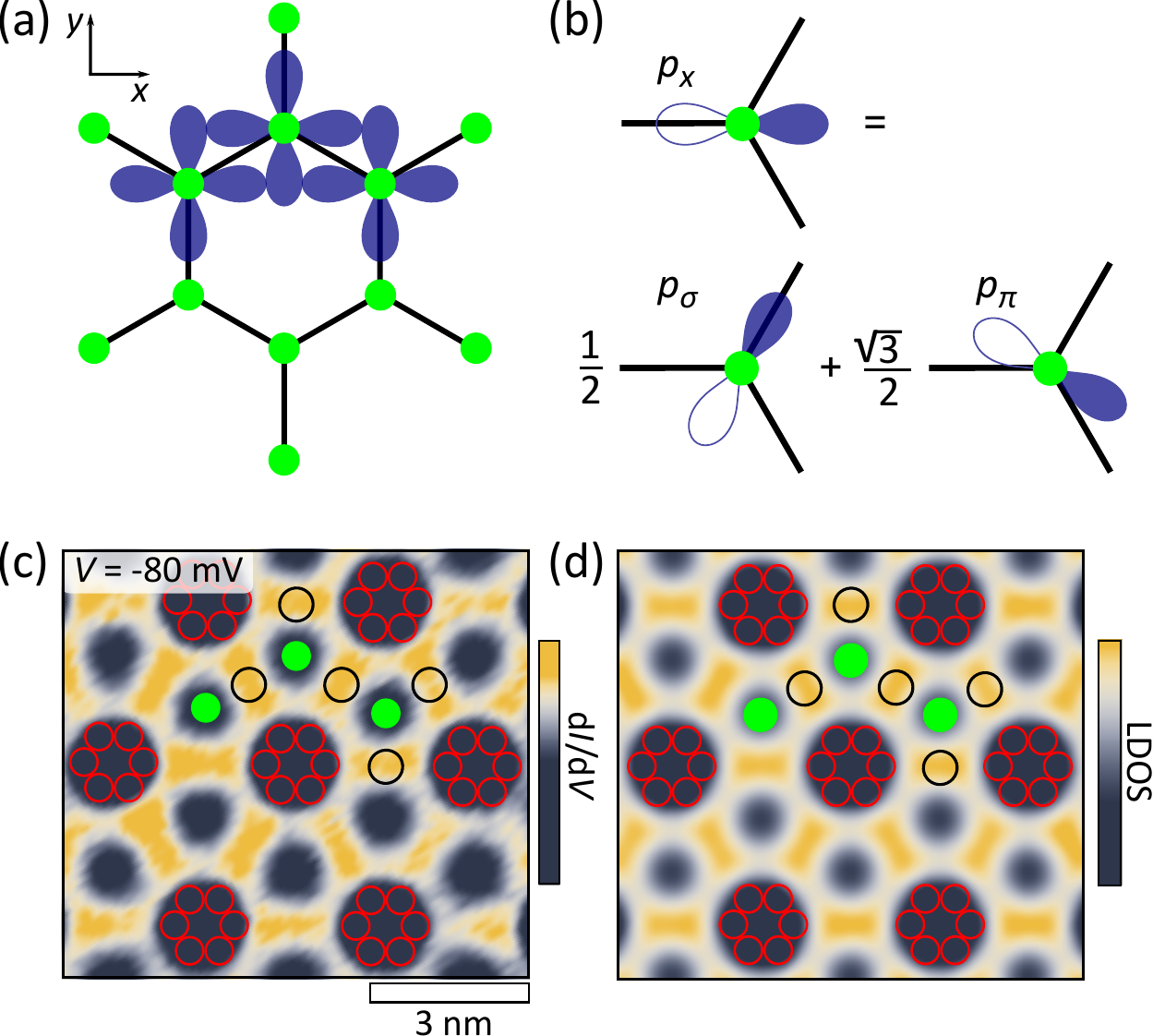}
\caption{(a) Schematic of $p$-orbitals (blue) in a honeycomb arrangement of artificial-atom sites (green). (b) Decomposition of a $p_x$-orbital into a $\sigma$- and $\pi$-bond. The rotation $\cos(60^\circ) =1/2$ and $\cos(-30^\circ) = \sqrt{3}/2$ give the prefactors for these rotations respectively. (c) Differential-conductance map acquired at $V= -80\,$mV above a honeycomb lattice, shown for a part of the lattice (see SM~\cite{Supp} for the entire lattice). Several artificial-atom sites (green), the $\sigma$-bond overlap of the $p$-like orbitals (black contours) and CO molecules (red contours) are indicated. The map was acquired at $-6\,$pm with respect to the setpoint $I_{\mathrm{set}} = 1\,$nA, $V_{\mathrm{set}} = 150\,$mV, with a bias modulation $V_{\mathrm{rms}} = 10\,$mV. 
(d) The corresponding LDOS map simulated using the muffin-tin model.}
\label{FIG4}
\end{figure}

We will now extend the approach to a lattice with a three-fold rotational symmetry about each artificial-atom site: the honeycomb lattice. Fig.~4a shows a schematic of a honeycomb lattice. To describe bonding between real atoms in such a geometry typically $sp^2$-hybridization is invoked. However, because the energy spacing between the $s$- and $p$-like orbitals can be made large, orbital hybridization does not necessarily occur. In this case, a well-established decomposition of the $p_x$-like and $p_y$-like orbitals into $\sigma$- and $\pi$-type components can be used to describe  the electron localization due to $p$-like orbitals in a triangular symmetry~\cite{Dresselhaus}. An example of a decomposition of a $p_x$-like orbital in a $\sigma$- and $\pi$-component is indicated in Fig. 4b. This leads to increased density of states in between the coupled artificial atoms and nodes at those sites. Fig. 4c shows a differential-conductance map of the as such realized honeycomb lattice acquired at $V = -80\,$mV. The artificial-atom sites are indicated in green and the CO molecules are encircled in red (distance between centers of CO clusters is $14 a$). The LDOS is highest at locations in between the artificial atoms (black contours), as expected from overlapping $p$-type orbitals with a $\sigma$-bond. The experimental observations are well reproduced by muffin-tin calculations (Fig.~4d). Note that since the decomposition method mentioned above can be used for any lattice symmetry, the $p$-like orbital description is expected to be generally applicable.

In conclusion, we demonstrated how to manipulate $p$-orbital bands in artificial electronic lattices with four-fold and three-fold rotational symmetry. In particular, we showed how the $p_x$- and $p_y$-orbitals can be tailored spatially and how the energy degeneracy of these states can be lifted, thus creating an analogue of the crystal-field splitting in these artificial lattices. We expect that the approach outlined here can be transferred to lattices created lithographically in semiconductors \cite{Tadjine}. The tunability of the geometries towards spherical structures \cite{Crommie} would facilitate adding a well-defined orbital angular momentum to the electrons, offering a platform for the investigation of $p$-orbital bands with spin-orbit coupling and their interaction with external fields.

\section*{Acknowledgements}
I.S., D.V. and C.M.S. acknowledge funding from NWO via grants 16PR3245, DDC 13, and 68047534, as well as an ERC Advanced Grant "FIRSTSTEP" 692691. A.A.K. acknowledges funding from NWO via VIDI grant 680-47-534.


\begin{thebibliography}{46}%
\makeatletter
\providecommand \@ifxundefined [1]{%
 \@ifx{#1\undefined}
}%
\providecommand \@ifnum [1]{%
 \ifnum #1\expandafter \@firstoftwo
 \else \expandafter \@secondoftwo
 \fi
}%
\providecommand \@ifx [1]{%
 \ifx #1\expandafter \@firstoftwo
 \else \expandafter \@secondoftwo
 \fi
}%
\providecommand \natexlab [1]{#1}%
\providecommand \enquote  [1]{``#1''}%
\providecommand \bibnamefont  [1]{#1}%
\providecommand \bibfnamefont [1]{#1}%
\providecommand \citenamefont [1]{#1}%
\providecommand \href@noop [0]{\@secondoftwo}%
\providecommand \href [0]{\begingroup \@sanitize@url \@href}%
\providecommand \@href[1]{\@@startlink{#1}\@@href}%
\providecommand \@@href[1]{\endgroup#1\@@endlink}%
\providecommand \@sanitize@url [0]{\catcode `\\12\catcode `\$12\catcode
  `\&12\catcode `\#12\catcode `\^12\catcode `\_12\catcode `\%12\relax}%
\providecommand \@@startlink[1]{}%
\providecommand \@@endlink[0]{}%
\providecommand \url  [0]{\begingroup\@sanitize@url \@url }%
\providecommand \@url [1]{\endgroup\@href {#1}{\urlprefix }}%
\providecommand \urlprefix  [0]{URL }%
\providecommand \Eprint [0]{\href }%
\providecommand \doibase [0]{http://dx.doi.org/}%
\providecommand \selectlanguage [0]{\@gobble}%
\providecommand \bibinfo  [0]{\@secondoftwo}%
\providecommand \bibfield  [0]{\@secondoftwo}%
\providecommand \translation [1]{[#1]}%
\providecommand \BibitemOpen [0]{}%
\providecommand \bibitemStop [0]{}%
\providecommand \bibitemNoStop [0]{.\EOS\space}%
\providecommand \EOS [0]{\spacefactor3000\relax}%
\providecommand \BibitemShut  [1]{\csname bibitem#1\endcsname}%
\let\auto@bib@innerbib\@empty
\bibitem [{\citenamefont {Tokura}\ and\ \citenamefont
  {Nagaosa}(2000)}]{Tokura}%
  \BibitemOpen
  \bibfield  {author} {\bibinfo {author} {\bibfnamefont {Y.}~\bibnamefont
  {Tokura}}\ and\ \bibinfo {author} {\bibfnamefont {N.}~\bibnamefont
  {Nagaosa}},\ }\href {\doibase 10.1126/science.288.5465.462} {\bibfield
  {journal} {\bibinfo  {journal} {Science}\ }\textbf {\bibinfo {volume}
  {288}},\ \bibinfo {pages} {462} (\bibinfo {year} {2000})}\BibitemShut
  {NoStop}%
\bibitem [{\citenamefont {Wu}\ \emph {et~al.}(2007)\citenamefont {Wu},
  \citenamefont {Bergman}, \citenamefont {Balents},\ and\ \citenamefont
  {Das~Sarma}}]{Wu2007}%
  \BibitemOpen
  \bibfield  {author} {\bibinfo {author} {\bibfnamefont {C.}~\bibnamefont
  {Wu}}, \bibinfo {author} {\bibfnamefont {D.}~\bibnamefont {Bergman}},
  \bibinfo {author} {\bibfnamefont {L.}~\bibnamefont {Balents}}, \ and\
  \bibinfo {author} {\bibfnamefont {S.}~\bibnamefont {Das~Sarma}},\ }\href
  {\doibase 10.1103/PhysRevLett.99.070401} {\bibfield  {journal} {\bibinfo
  {journal} {Phys. Rev. Lett.}\ }\textbf {\bibinfo {volume} {99}},\ \bibinfo
  {pages} {070401} (\bibinfo {year} {2007})}\BibitemShut {NoStop}%
\bibitem [{\citenamefont {Wu}\ and\ \citenamefont {Das~Sarma}(2008)}]{Wu2008}%
  \BibitemOpen
  \bibfield  {author} {\bibinfo {author} {\bibfnamefont {C.}~\bibnamefont
  {Wu}}\ and\ \bibinfo {author} {\bibfnamefont {S.}~\bibnamefont {Das~Sarma}},\
  }\href {\doibase 10.1103/PhysRevB.77.235107} {\bibfield  {journal} {\bibinfo
  {journal} {Phys. Rev. B}\ }\textbf {\bibinfo {volume} {77}},\ \bibinfo
  {pages} {235107} (\bibinfo {year} {2008})}\BibitemShut {NoStop}%
\bibitem [{\citenamefont {Beugeling}\ \emph {et~al.}(2015)\citenamefont
  {Beugeling}, \citenamefont {Kalesaki}, \citenamefont {Delerue}, \citenamefont
  {Niquet}, \citenamefont {Vanmaekelbergh},\ and\ \citenamefont
  {Smith}}]{Beugeling}%
  \BibitemOpen
  \bibfield  {author} {\bibinfo {author} {\bibfnamefont {W.}~\bibnamefont
  {Beugeling}}, \bibinfo {author} {\bibfnamefont {E.}~\bibnamefont {Kalesaki}},
  \bibinfo {author} {\bibfnamefont {C.}~\bibnamefont {Delerue}}, \bibinfo
  {author} {\bibfnamefont {Y.-M.}\ \bibnamefont {Niquet}}, \bibinfo {author}
  {\bibfnamefont {D.}~\bibnamefont {Vanmaekelbergh}}, \ and\ \bibinfo {author}
  {\bibfnamefont {C.~M.}\ \bibnamefont {Smith}},\ }\href@noop {} {\bibfield
  {journal} {\bibinfo  {journal} {Nature Communications}\ }\textbf {\bibinfo
  {volume} {6}},\ \bibinfo {pages} {6316} (\bibinfo {year} {2015})}\BibitemShut
  {NoStop}%
\bibitem [{\citenamefont {Wirth}\ \emph {et~al.}(2011)\citenamefont {Wirth},
  \citenamefont {\"Olschl\"ager},\ and\ \citenamefont {Hemmerich}}]{Wirth}%
  \BibitemOpen
  \bibfield  {author} {\bibinfo {author} {\bibfnamefont {G.}~\bibnamefont
  {Wirth}}, \bibinfo {author} {\bibfnamefont {M.}~\bibnamefont
  {\"Olschl\"ager}}, \ and\ \bibinfo {author} {\bibfnamefont {A.}~\bibnamefont
  {Hemmerich}},\ }\href {\doibase 10.1038/nphys1857} {\bibfield  {journal}
  {\bibinfo  {journal} {Nature Physics}\ }\textbf {\bibinfo {volume} {7}},\
  \bibinfo {pages} {147} (\bibinfo {year} {2011})}\BibitemShut {NoStop}%
\bibitem [{\citenamefont {\"Olschl\"ager}\ \emph {et~al.}(2012)\citenamefont
  {\"Olschl\"ager}, \citenamefont {Wirth}, \citenamefont {Kock},\ and\
  \citenamefont {Hemmerich}}]{Oelschlaeger}%
  \BibitemOpen
  \bibfield  {author} {\bibinfo {author} {\bibfnamefont {M.}~\bibnamefont
  {\"Olschl\"ager}}, \bibinfo {author} {\bibfnamefont {G.}~\bibnamefont
  {Wirth}}, \bibinfo {author} {\bibfnamefont {T.}~\bibnamefont {Kock}}, \ and\
  \bibinfo {author} {\bibfnamefont {A.}~\bibnamefont {Hemmerich}},\ }\href
  {\doibase 10.1103/PhysRevLett.108.075302} {\bibfield  {journal} {\bibinfo
  {journal} {Phys. Rev. Lett.}\ }\textbf {\bibinfo {volume} {108}},\ \bibinfo
  {pages} {075302} (\bibinfo {year} {2012})}\BibitemShut {NoStop}%
\bibitem [{\citenamefont {M\"uller}\ \emph {et~al.}(2007)\citenamefont
  {M\"uller}, \citenamefont {F\"olling}, \citenamefont {Widera},\ and\
  \citenamefont {Bloch}}]{Mueller}%
  \BibitemOpen
  \bibfield  {author} {\bibinfo {author} {\bibfnamefont {T.}~\bibnamefont
  {M\"uller}}, \bibinfo {author} {\bibfnamefont {S.}~\bibnamefont {F\"olling}},
  \bibinfo {author} {\bibfnamefont {A.}~\bibnamefont {Widera}}, \ and\ \bibinfo
  {author} {\bibfnamefont {I.}~\bibnamefont {Bloch}},\ }\href {\doibase
  10.1103/PhysRevLett.99.200405} {\bibfield  {journal} {\bibinfo  {journal}
  {Phys. Rev. Lett.}\ }\textbf {\bibinfo {volume} {99}},\ \bibinfo {pages}
  {200405} (\bibinfo {year} {2007})}\BibitemShut {NoStop}%
\bibitem [{\citenamefont {\"Olschl\"ager}\ \emph {et~al.}(2013)\citenamefont
  {\"Olschl\"ager}, \citenamefont {Kock}, \citenamefont {Wirth}, \citenamefont
  {Ewerbeck}, \citenamefont {Smith},\ and\ \citenamefont
  {Hemmerich}}]{SmithHemmerich}%
  \BibitemOpen
  \bibfield  {author} {\bibinfo {author} {\bibfnamefont {M.}~\bibnamefont
  {\"Olschl\"ager}}, \bibinfo {author} {\bibfnamefont {T.}~\bibnamefont
  {Kock}}, \bibinfo {author} {\bibfnamefont {G.}~\bibnamefont {Wirth}},
  \bibinfo {author} {\bibfnamefont {A.}~\bibnamefont {Ewerbeck}}, \bibinfo
  {author} {\bibfnamefont {C.~M.}\ \bibnamefont {Smith}}, \ and\ \bibinfo
  {author} {\bibfnamefont {A.}~\bibnamefont {Hemmerich}},\ }\href@noop {}
  {\bibfield  {journal} {\bibinfo  {journal} {New Journal of Physics}\ }\textbf
  {\bibinfo {volume} {15}},\ \bibinfo {pages} {083041} (\bibinfo {year}
  {2013})}\BibitemShut {NoStop}%
\bibitem [{\citenamefont {Kirtley}\ \emph {et~al.}(1995)\citenamefont
  {Kirtley}, \citenamefont {Tsuei}, \citenamefont {Sun}, \citenamefont {Chi},
  \citenamefont {Yu-Jahnes}, \citenamefont {Gupta}, \citenamefont {Rupp},\ and\
  \citenamefont {Ketchen}}]{Kirtley}%
  \BibitemOpen
  \bibfield  {author} {\bibinfo {author} {\bibfnamefont {J.}~\bibnamefont
  {Kirtley}}, \bibinfo {author} {\bibfnamefont {C.}~\bibnamefont {Tsuei}},
  \bibinfo {author} {\bibfnamefont {J.}~\bibnamefont {Sun}}, \bibinfo {author}
  {\bibfnamefont {C.}~\bibnamefont {Chi}}, \bibinfo {author} {\bibfnamefont
  {L.~S.}\ \bibnamefont {Yu-Jahnes}}, \bibinfo {author} {\bibfnamefont
  {A.}~\bibnamefont {Gupta}}, \bibinfo {author} {\bibfnamefont
  {M.}~\bibnamefont {Rupp}}, \ and\ \bibinfo {author} {\bibfnamefont
  {M.}~\bibnamefont {Ketchen}},\ }\href@noop {} {\bibfield  {journal} {\bibinfo
   {journal} {Nature}\ }\textbf {\bibinfo {volume} {373}},\ \bibinfo {pages}
  {225} (\bibinfo {year} {1995})}\BibitemShut {NoStop}%
\bibitem [{\citenamefont {Maeno}\ \emph {et~al.}(1994)\citenamefont {Maeno},
  \citenamefont {Hashimoto}, \citenamefont {Yoshida}, \citenamefont
  {Nishizaki}, \citenamefont {Fujita}, \citenamefont {Bednorz},\ and\
  \citenamefont {Lichtenberg}}]{Maeno}%
  \BibitemOpen
  \bibfield  {author} {\bibinfo {author} {\bibfnamefont {Y.}~\bibnamefont
  {Maeno}}, \bibinfo {author} {\bibfnamefont {H.}~\bibnamefont {Hashimoto}},
  \bibinfo {author} {\bibfnamefont {K.}~\bibnamefont {Yoshida}}, \bibinfo
  {author} {\bibfnamefont {S.}~\bibnamefont {Nishizaki}}, \bibinfo {author}
  {\bibfnamefont {T.}~\bibnamefont {Fujita}}, \bibinfo {author} {\bibfnamefont
  {J.}~\bibnamefont {Bednorz}}, \ and\ \bibinfo {author} {\bibfnamefont
  {F.}~\bibnamefont {Lichtenberg}},\ }\href@noop {} {\bibfield  {journal}
  {\bibinfo  {journal} {Nature}\ }\textbf {\bibinfo {volume} {372}},\ \bibinfo
  {pages} {532} (\bibinfo {year} {1994})}\BibitemShut {NoStop}%
\bibitem [{\citenamefont {Mackenzie}\ \emph {et~al.}(2017)\citenamefont
  {Mackenzie}, \citenamefont {Scaffidi}, \citenamefont {Hicks},\ and\
  \citenamefont {Maeno}}]{Mackenzie}%
  \BibitemOpen
  \bibfield  {author} {\bibinfo {author} {\bibfnamefont {A.~P.}\ \bibnamefont
  {Mackenzie}}, \bibinfo {author} {\bibfnamefont {T.}~\bibnamefont {Scaffidi}},
  \bibinfo {author} {\bibfnamefont {C.~W.}\ \bibnamefont {Hicks}}, \ and\
  \bibinfo {author} {\bibfnamefont {Y.}~\bibnamefont {Maeno}},\ }\href@noop {}
  {\bibfield  {journal} {\bibinfo  {journal} {npj Quantum Materials}\ }\textbf
  {\bibinfo {volume} {2}},\ \bibinfo {pages} {40} (\bibinfo {year}
  {2017})}\BibitemShut {NoStop}%
\bibitem [{\citenamefont {Kvorning}\ \emph {et~al.}(2018)\citenamefont
  {Kvorning}, \citenamefont {Hansson}, \citenamefont {Quelle},\ and\
  \citenamefont {Smith}}]{Kvorning}%
  \BibitemOpen
  \bibfield  {author} {\bibinfo {author} {\bibfnamefont {T.}~\bibnamefont
  {Kvorning}}, \bibinfo {author} {\bibfnamefont {T.~H.}\ \bibnamefont
  {Hansson}}, \bibinfo {author} {\bibfnamefont {A.}~\bibnamefont {Quelle}}, \
  and\ \bibinfo {author} {\bibfnamefont {C.~M.}\ \bibnamefont {Smith}},\
  }\href@noop {} {\bibfield  {journal} {\bibinfo  {journal} {Phys. Rev. Lett.}\
  }\textbf {\bibinfo {volume} {120}},\ \bibinfo {pages} {217002} (\bibinfo
  {year} {2018})}\BibitemShut {NoStop}%
\bibitem [{\citenamefont {Kock}\ \emph {et~al.}(2016)\citenamefont {Kock},
  \citenamefont {Hippler}, \citenamefont {Ewerbeck},\ and\ \citenamefont
  {Hemmerich}}]{Kock}%
  \BibitemOpen
  \bibfield  {author} {\bibinfo {author} {\bibfnamefont {T.}~\bibnamefont
  {Kock}}, \bibinfo {author} {\bibfnamefont {C.}~\bibnamefont {Hippler}},
  \bibinfo {author} {\bibfnamefont {A.}~\bibnamefont {Ewerbeck}}, \ and\
  \bibinfo {author} {\bibfnamefont {A.}~\bibnamefont {Hemmerich}},\ }\href@noop
  {} {\bibfield  {journal} {\bibinfo  {journal} {Journal of Physics B: Atomic,
  Molecular and Optical Physics}\ }\textbf {\bibinfo {volume} {49}},\ \bibinfo
  {pages} {042001} (\bibinfo {year} {2016})}\BibitemShut {NoStop}%
\bibitem [{\citenamefont {Li}\ and\ \citenamefont {Liu}(2016)}]{Li}%
  \BibitemOpen
  \bibfield  {author} {\bibinfo {author} {\bibfnamefont {X.}~\bibnamefont
  {Li}}\ and\ \bibinfo {author} {\bibfnamefont {W.~V.}\ \bibnamefont {Liu}},\
  }\href@noop {} {\bibfield  {journal} {\bibinfo  {journal} {Reports on
  Progress in Physics}\ }\textbf {\bibinfo {volume} {79}},\ \bibinfo {pages}
  {116401} (\bibinfo {year} {2016})}\BibitemShut {NoStop}%
\bibitem [{\citenamefont {Lewenstein}\ and\ \citenamefont
  {Liu}(2011)}]{Lewenstein}%
  \BibitemOpen
  \bibfield  {author} {\bibinfo {author} {\bibfnamefont {M.}~\bibnamefont
  {Lewenstein}}\ and\ \bibinfo {author} {\bibfnamefont {W.~V.}\ \bibnamefont
  {Liu}},\ }\href@noop {} {\bibfield  {journal} {\bibinfo  {journal} {Nature
  Physics}\ }\textbf {\bibinfo {volume} {7}},\ \bibinfo {pages} {101} (\bibinfo
  {year} {2011})}\BibitemShut {NoStop}%
\bibitem [{\citenamefont {Mili\ifmmode \acute{c}\else
  \'{c}\fi{}evi\ifmmode~\acute{c}\else \'{c}\fi{}}\ \emph
  {et~al.}(2017)\citenamefont {Mili\ifmmode \acute{c}\else
  \'{c}\fi{}evi\ifmmode~\acute{c}\else \'{c}\fi{}}, \citenamefont {Ozawa},
  \citenamefont {Montambaux}, \citenamefont {Carusotto}, \citenamefont
  {Galopin}, \citenamefont {Lema\^{\i}tre}, \citenamefont {Le~Gratiet},
  \citenamefont {Sagnes}, \citenamefont {Bloch},\ and\ \citenamefont
  {Amo}}]{Milicevic}%
  \BibitemOpen
  \bibfield  {author} {\bibinfo {author} {\bibfnamefont {M.}~\bibnamefont
  {Mili\ifmmode \acute{c}\else \'{c}\fi{}evi\ifmmode~\acute{c}\else
  \'{c}\fi{}}}, \bibinfo {author} {\bibfnamefont {T.}~\bibnamefont {Ozawa}},
  \bibinfo {author} {\bibfnamefont {G.}~\bibnamefont {Montambaux}}, \bibinfo
  {author} {\bibfnamefont {I.}~\bibnamefont {Carusotto}}, \bibinfo {author}
  {\bibfnamefont {E.}~\bibnamefont {Galopin}}, \bibinfo {author} {\bibfnamefont
  {A.}~\bibnamefont {Lema\^{\i}tre}}, \bibinfo {author} {\bibfnamefont
  {L.}~\bibnamefont {Le~Gratiet}}, \bibinfo {author} {\bibfnamefont
  {I.}~\bibnamefont {Sagnes}}, \bibinfo {author} {\bibfnamefont
  {J.}~\bibnamefont {Bloch}}, \ and\ \bibinfo {author} {\bibfnamefont
  {A.}~\bibnamefont {Amo}},\ }\href {\doibase 10.1103/PhysRevLett.118.107403}
  {\bibfield  {journal} {\bibinfo  {journal} {Phys. Rev. Lett.}\ }\textbf
  {\bibinfo {volume} {118}},\ \bibinfo {pages} {107403} (\bibinfo {year}
  {2017})}\BibitemShut {NoStop}%
\bibitem [{\citenamefont {Cantillano}\ \emph {et~al.}(2018)\citenamefont
  {Cantillano}, \citenamefont {Mukherjee}, \citenamefont {Morales-Inostroza},
  \citenamefont {Real}, \citenamefont {Cáceres-Aravena}, \citenamefont
  {Hermann-Avigliano}, \citenamefont {Thomson},\ and\ \citenamefont
  {Vicencio}}]{Cantillano}%
  \BibitemOpen
  \bibfield  {author} {\bibinfo {author} {\bibfnamefont {C.}~\bibnamefont
  {Cantillano}}, \bibinfo {author} {\bibfnamefont {S.}~\bibnamefont
  {Mukherjee}}, \bibinfo {author} {\bibfnamefont {L.}~\bibnamefont
  {Morales-Inostroza}}, \bibinfo {author} {\bibfnamefont {B.}~\bibnamefont
  {Real}}, \bibinfo {author} {\bibfnamefont {G.}~\bibnamefont
  {Cáceres-Aravena}}, \bibinfo {author} {\bibfnamefont {C.}~\bibnamefont
  {Hermann-Avigliano}}, \bibinfo {author} {\bibfnamefont {R.~R.}\ \bibnamefont
  {Thomson}}, \ and\ \bibinfo {author} {\bibfnamefont {R.~A.}\ \bibnamefont
  {Vicencio}},\ }\href@noop {} {\bibfield  {journal} {\bibinfo  {journal} {New
  Journal of Physics}\ }\textbf {\bibinfo {volume} {20}},\ \bibinfo {pages}
  {033028} (\bibinfo {year} {2018})}\BibitemShut {NoStop}%
\bibitem [{\citenamefont {Jacqmin}\ \emph {et~al.}(2014)\citenamefont
  {Jacqmin}, \citenamefont {Carusotto}, \citenamefont {Sagnes}, \citenamefont
  {Abbarchi}, \citenamefont {Solnyshkov}, \citenamefont {Malpuech},
  \citenamefont {Galopin}, \citenamefont {Lema\^{\i}tre}, \citenamefont
  {Bloch},\ and\ \citenamefont {Amo}}]{Jacqmin}%
  \BibitemOpen
  \bibfield  {author} {\bibinfo {author} {\bibfnamefont {T.}~\bibnamefont
  {Jacqmin}}, \bibinfo {author} {\bibfnamefont {I.}~\bibnamefont {Carusotto}},
  \bibinfo {author} {\bibfnamefont {I.}~\bibnamefont {Sagnes}}, \bibinfo
  {author} {\bibfnamefont {M.}~\bibnamefont {Abbarchi}}, \bibinfo {author}
  {\bibfnamefont {D.~D.}\ \bibnamefont {Solnyshkov}}, \bibinfo {author}
  {\bibfnamefont {G.}~\bibnamefont {Malpuech}}, \bibinfo {author}
  {\bibfnamefont {E.}~\bibnamefont {Galopin}}, \bibinfo {author} {\bibfnamefont
  {A.}~\bibnamefont {Lema\^{\i}tre}}, \bibinfo {author} {\bibfnamefont
  {J.}~\bibnamefont {Bloch}}, \ and\ \bibinfo {author} {\bibfnamefont
  {A.}~\bibnamefont {Amo}},\ }\href {\doibase 10.1103/PhysRevLett.112.116402}
  {\bibfield  {journal} {\bibinfo  {journal} {Phys. Rev. Lett.}\ }\textbf
  {\bibinfo {volume} {112}},\ \bibinfo {pages} {116402} (\bibinfo {year}
  {2014})}\BibitemShut {NoStop}%
\bibitem [{\citenamefont {Klembt}\ \emph {et~al.}(2017)\citenamefont {Klembt},
  \citenamefont {Harder}, \citenamefont {Egorov}, \citenamefont {Winkler},
  \citenamefont {Suchomel}, \citenamefont {Beierlein}, \citenamefont
  {Emmerling}, \citenamefont {Schneider},\ and\ \citenamefont
  {H\"{o}fling}}]{Klembt}%
  \BibitemOpen
  \bibfield  {author} {\bibinfo {author} {\bibfnamefont {S.}~\bibnamefont
  {Klembt}}, \bibinfo {author} {\bibfnamefont {T.~H.}\ \bibnamefont {Harder}},
  \bibinfo {author} {\bibfnamefont {O.~A.}\ \bibnamefont {Egorov}}, \bibinfo
  {author} {\bibfnamefont {K.}~\bibnamefont {Winkler}}, \bibinfo {author}
  {\bibfnamefont {H.}~\bibnamefont {Suchomel}}, \bibinfo {author}
  {\bibfnamefont {J.}~\bibnamefont {Beierlein}}, \bibinfo {author}
  {\bibfnamefont {M.}~\bibnamefont {Emmerling}}, \bibinfo {author}
  {\bibfnamefont {C.}~\bibnamefont {Schneider}}, \ and\ \bibinfo {author}
  {\bibfnamefont {S.}~\bibnamefont {H\"{o}fling}},\ }\href@noop {} {\bibfield
  {journal} {\bibinfo  {journal} {Applied Physics Letters}\ }\textbf {\bibinfo
  {volume} {111}},\ \bibinfo {pages} {231102} (\bibinfo {year}
  {2017})}\BibitemShut {NoStop}%
\bibitem [{\citenamefont {Whittaker}\ \emph {et~al.}(2018)\citenamefont
  {Whittaker}, \citenamefont {Cancellieri}, \citenamefont {Walker},
  \citenamefont {Gulevich}, \citenamefont {Schomerus}, \citenamefont
  {Vaitiekus}, \citenamefont {Royall}, \citenamefont {Whittaker}, \citenamefont
  {Clarke}, \citenamefont {Iorsh}, \citenamefont {Shelykh}, \citenamefont
  {Skolnick},\ and\ \citenamefont {Krizhanovskii}}]{Whittaker}%
  \BibitemOpen
  \bibfield  {author} {\bibinfo {author} {\bibfnamefont {C.~E.}\ \bibnamefont
  {Whittaker}}, \bibinfo {author} {\bibfnamefont {E.}~\bibnamefont
  {Cancellieri}}, \bibinfo {author} {\bibfnamefont {P.~M.}\ \bibnamefont
  {Walker}}, \bibinfo {author} {\bibfnamefont {D.~R.}\ \bibnamefont
  {Gulevich}}, \bibinfo {author} {\bibfnamefont {H.}~\bibnamefont {Schomerus}},
  \bibinfo {author} {\bibfnamefont {D.}~\bibnamefont {Vaitiekus}}, \bibinfo
  {author} {\bibfnamefont {B.}~\bibnamefont {Royall}}, \bibinfo {author}
  {\bibfnamefont {D.~M.}\ \bibnamefont {Whittaker}}, \bibinfo {author}
  {\bibfnamefont {E.}~\bibnamefont {Clarke}}, \bibinfo {author} {\bibfnamefont
  {I.~V.}\ \bibnamefont {Iorsh}}, \bibinfo {author} {\bibfnamefont {I.~A.}\
  \bibnamefont {Shelykh}}, \bibinfo {author} {\bibfnamefont {M.~S.}\
  \bibnamefont {Skolnick}}, \ and\ \bibinfo {author} {\bibfnamefont {D.~N.}\
  \bibnamefont {Krizhanovskii}},\ }\href@noop {} {\bibfield  {journal}
  {\bibinfo  {journal} {Phys. Rev. Lett.}\ }\textbf {\bibinfo {volume} {120}},\
  \bibinfo {pages} {097401} (\bibinfo {year} {2018})}\BibitemShut {NoStop}%
\bibitem [{\citenamefont {{Nilius}}\ \emph {et~al.}(2002)\citenamefont
  {{Nilius}}, \citenamefont {{Wallis}},\ and\ \citenamefont {{Ho}}}]{Nilius}%
  \BibitemOpen
  \bibfield  {author} {\bibinfo {author} {\bibfnamefont {N.}~\bibnamefont
  {{Nilius}}}, \bibinfo {author} {\bibfnamefont {T.~M.}\ \bibnamefont
  {{Wallis}}}, \ and\ \bibinfo {author} {\bibfnamefont {W.}~\bibnamefont
  {{Ho}}},\ }\href@noop {} {\bibfield  {journal} {\bibinfo  {journal}
  {Science}\ }\textbf {\bibinfo {volume} {297}},\ \bibinfo {pages} {1853}
  (\bibinfo {year} {2002})}\BibitemShut {NoStop}%
\bibitem [{\citenamefont {{Nilius}}\ \emph {et~al.}(2014)\citenamefont
  {{Nilius}}, \citenamefont {{Wallis}}, \citenamefont {{Persson}},\ and\
  \citenamefont {{Ho}}}]{Nilius2}%
  \BibitemOpen
  \bibfield  {author} {\bibinfo {author} {\bibfnamefont {N.}~\bibnamefont
  {{Nilius}}}, \bibinfo {author} {\bibfnamefont {T.~M.}\ \bibnamefont
  {{Wallis}}}, \bibinfo {author} {\bibfnamefont {M.}~\bibnamefont {{Persson}}},
  \ and\ \bibinfo {author} {\bibfnamefont {W.}~\bibnamefont {{Ho}}},\
  }\href@noop {} {\bibfield  {journal} {\bibinfo  {journal} {The Journal of
  Physical Chemistry C}\ }\textbf {\bibinfo {volume} {118}},\ \bibinfo {pages}
  {29001} (\bibinfo {year} {2014})}\BibitemShut {NoStop}%
\bibitem [{\citenamefont {{F\"{o}lsch}}\ \emph {et~al.}(2014)\citenamefont
  {{F\"{o}lsch}}, \citenamefont {{Hyldgaard}}, \citenamefont {{Koch}},\ and\
  \citenamefont {{Ploog}}}]{Folsch}%
  \BibitemOpen
  \bibfield  {author} {\bibinfo {author} {\bibfnamefont {S.}~\bibnamefont
  {{F\"{o}lsch}}}, \bibinfo {author} {\bibfnamefont {P.}~\bibnamefont
  {{Hyldgaard}}}, \bibinfo {author} {\bibfnamefont {R.}~\bibnamefont {{Koch}}},
  \ and\ \bibinfo {author} {\bibfnamefont {K.}~\bibnamefont {{Ploog}}},\
  }\href@noop {} {\bibfield  {journal} {\bibinfo  {journal} {The Journal of
  Physical Chemistry C}\ }\textbf {\bibinfo {volume} {118}},\ \bibinfo {pages}
  {29001} (\bibinfo {year} {2014})}\BibitemShut {NoStop}%
\bibitem [{\citenamefont {Hirjibehedin}\ \emph {et~al.}(2006)\citenamefont
  {Hirjibehedin}, \citenamefont {Lutz},\ and\ \citenamefont
  {Heinrich}}]{Hirjibehedin2006}%
  \BibitemOpen
  \bibfield  {author} {\bibinfo {author} {\bibfnamefont {C.~F.}\ \bibnamefont
  {Hirjibehedin}}, \bibinfo {author} {\bibfnamefont {C.~P.}\ \bibnamefont
  {Lutz}}, \ and\ \bibinfo {author} {\bibfnamefont {A.~J.}\ \bibnamefont
  {Heinrich}},\ }\href@noop {} {\bibfield  {journal} {\bibinfo  {journal}
  {Science}\ }\textbf {\bibinfo {volume} {312}},\ \bibinfo {pages} {1021}
  (\bibinfo {year} {2006})}\BibitemShut {NoStop}%
\bibitem [{\citenamefont {Khajetoorians}\ \emph {et~al.}(2012)\citenamefont
  {Khajetoorians}, \citenamefont {Wiebe}, \citenamefont {Chilian},
  \citenamefont {Lounis}, \citenamefont {Bl\"{u}gel},\ and\ \citenamefont
  {Wiesendanger}}]{Khajetoorians2012}%
  \BibitemOpen
  \bibfield  {author} {\bibinfo {author} {\bibfnamefont {A.~A.}\ \bibnamefont
  {Khajetoorians}}, \bibinfo {author} {\bibfnamefont {J.}~\bibnamefont
  {Wiebe}}, \bibinfo {author} {\bibfnamefont {B.}~\bibnamefont {Chilian}},
  \bibinfo {author} {\bibfnamefont {S.}~\bibnamefont {Lounis}}, \bibinfo
  {author} {\bibfnamefont {S.}~\bibnamefont {Bl\"{u}gel}}, \ and\ \bibinfo
  {author} {\bibfnamefont {R.}~\bibnamefont {Wiesendanger}},\ }\href@noop {}
  {\bibfield  {journal} {\bibinfo  {journal} {Nature Physics}\ }\textbf
  {\bibinfo {volume} {8}},\ \bibinfo {pages} {497} (\bibinfo {year}
  {2012})}\BibitemShut {NoStop}%
\bibitem [{\citenamefont {Kamlapure}\ \emph {et~al.}(2018)\citenamefont
  {Kamlapure}, \citenamefont {Cornils}, \citenamefont {Wiebe},\ and\
  \citenamefont {Wiesendanger}}]{Kamlapure}%
  \BibitemOpen
  \bibfield  {author} {\bibinfo {author} {\bibfnamefont {A.}~\bibnamefont
  {Kamlapure}}, \bibinfo {author} {\bibfnamefont {L.}~\bibnamefont {Cornils}},
  \bibinfo {author} {\bibfnamefont {J.}~\bibnamefont {Wiebe}}, \ and\ \bibinfo
  {author} {\bibfnamefont {R.}~\bibnamefont {Wiesendanger}},\ }\href@noop {}
  {\bibfield  {journal} {\bibinfo  {journal} {Nature Communications}\ }\textbf
  {\bibinfo {volume} {9}},\ \bibinfo {pages} {3253} (\bibinfo {year}
  {2018})}\BibitemShut {NoStop}%
\bibitem [{\citenamefont {Gomes}\ \emph {et~al.}(2012)\citenamefont {Gomes},
  \citenamefont {Mar}, \citenamefont {Ko}, \citenamefont {Guinea},\ and\
  \citenamefont {Manoharan}}]{Manoharan}%
  \BibitemOpen
  \bibfield  {author} {\bibinfo {author} {\bibfnamefont {K.~K.}\ \bibnamefont
  {Gomes}}, \bibinfo {author} {\bibfnamefont {W.}~\bibnamefont {Mar}}, \bibinfo
  {author} {\bibfnamefont {W.}~\bibnamefont {Ko}}, \bibinfo {author}
  {\bibfnamefont {F.}~\bibnamefont {Guinea}}, \ and\ \bibinfo {author}
  {\bibfnamefont {H.~C.}\ \bibnamefont {Manoharan}},\ }\href@noop {} {\bibfield
   {journal} {\bibinfo  {journal} {Nature}\ }\textbf {\bibinfo {volume}
  {483}},\ \bibinfo {pages} {306} (\bibinfo {year} {2012})}\BibitemShut
  {NoStop}%
\bibitem [{\citenamefont {{Drost}}\ \emph {et~al.}(2017)\citenamefont
  {{Drost}}, \citenamefont {{Ojanen}}, \citenamefont {{Harju}},\ and\
  \citenamefont {{Liljeroth}}}]{Drost2017}%
  \BibitemOpen
  \bibfield  {author} {\bibinfo {author} {\bibfnamefont {R.}~\bibnamefont
  {{Drost}}}, \bibinfo {author} {\bibfnamefont {T.}~\bibnamefont {{Ojanen}}},
  \bibinfo {author} {\bibfnamefont {A.}~\bibnamefont {{Harju}}}, \ and\
  \bibinfo {author} {\bibfnamefont {P.}~\bibnamefont {{Liljeroth}}},\
  }\href@noop {} {\bibfield  {journal} {\bibinfo  {journal} {Nature Physics}\
  }\textbf {\bibinfo {volume} {13}},\ \bibinfo {pages} {668} (\bibinfo {year}
  {2017})}\BibitemShut {NoStop}%
\bibitem [{\citenamefont {{Slot}}\ \emph {et~al.}(2017)\citenamefont {{Slot}},
  \citenamefont {{Gardenier}}, \citenamefont {{Jacobse}}, \citenamefont {{van
  Miert}}, \citenamefont {{Kempkes}}, \citenamefont {{Zevenhuizen}},
  \citenamefont {{Smith}}, \citenamefont {{Vanmaekelbergh}},\ and\
  \citenamefont {{Swart}}}]{Slot2017}%
  \BibitemOpen
  \bibfield  {author} {\bibinfo {author} {\bibfnamefont {M.~R.}\ \bibnamefont
  {{Slot}}}, \bibinfo {author} {\bibfnamefont {T.~S.}\ \bibnamefont
  {{Gardenier}}}, \bibinfo {author} {\bibfnamefont {P.~H.}\ \bibnamefont
  {{Jacobse}}}, \bibinfo {author} {\bibfnamefont {G.~C.~P.}\ \bibnamefont {{van
  Miert}}}, \bibinfo {author} {\bibfnamefont {S.~N.}\ \bibnamefont
  {{Kempkes}}}, \bibinfo {author} {\bibfnamefont {S.~J.~M.}\ \bibnamefont
  {{Zevenhuizen}}}, \bibinfo {author} {\bibfnamefont {C.~M.}\ \bibnamefont
  {{Smith}}}, \bibinfo {author} {\bibfnamefont {D.}~\bibnamefont
  {{Vanmaekelbergh}}}, \ and\ \bibinfo {author} {\bibfnamefont
  {I.}~\bibnamefont {{Swart}}},\ }\href {\doibase 10.1038/nphys4105} {\bibfield
   {journal} {\bibinfo  {journal} {Nature Physics}\ }\textbf {\bibinfo {volume}
  {13}},\ \bibinfo {pages} {672} (\bibinfo {year} {2017})}\BibitemShut
  {NoStop}%
\bibitem [{\citenamefont {Girovsky}\ \emph {et~al.}(2017)\citenamefont
  {Girovsky}, \citenamefont {Lado}, \citenamefont {Kalff}, \citenamefont
  {Fahrenfort}, \citenamefont {Peters}, \citenamefont {Fernández-Rossier},\
  and\ \citenamefont {Otte}}]{Otte}%
  \BibitemOpen
  \bibfield  {author} {\bibinfo {author} {\bibfnamefont {J.}~\bibnamefont
  {Girovsky}}, \bibinfo {author} {\bibfnamefont {J.~L.}\ \bibnamefont {Lado}},
  \bibinfo {author} {\bibfnamefont {F.~E.}\ \bibnamefont {Kalff}}, \bibinfo
  {author} {\bibfnamefont {E.}~\bibnamefont {Fahrenfort}}, \bibinfo {author}
  {\bibfnamefont {L.~J. J.~M.}\ \bibnamefont {Peters}}, \bibinfo {author}
  {\bibfnamefont {J.}~\bibnamefont {Fernández-Rossier}}, \ and\ \bibinfo
  {author} {\bibfnamefont {A.~F.}\ \bibnamefont {Otte}},\ }\href@noop {}
  {\bibfield  {journal} {\bibinfo  {journal} {SciPost Phys.}\ }\textbf
  {\bibinfo {volume} {2}},\ \bibinfo {pages} {020} (\bibinfo {year}
  {2017})}\BibitemShut {NoStop}%
\bibitem [{\citenamefont {{Collins}}\ \emph {et~al.}(2017)\citenamefont
  {{Collins}}, \citenamefont {{Witte}}, \citenamefont {{Silverman}},
  \citenamefont {{Green}},\ and\ \citenamefont {{Gomes}}}]{Collins2017}%
  \BibitemOpen
  \bibfield  {author} {\bibinfo {author} {\bibfnamefont {L.~C.}\ \bibnamefont
  {{Collins}}}, \bibinfo {author} {\bibfnamefont {T.~G.}\ \bibnamefont
  {{Witte}}}, \bibinfo {author} {\bibfnamefont {R.}~\bibnamefont
  {{Silverman}}}, \bibinfo {author} {\bibfnamefont {D.~B.}\ \bibnamefont
  {{Green}}}, \ and\ \bibinfo {author} {\bibfnamefont {K.~K.}\ \bibnamefont
  {{Gomes}}},\ }\href {\doibase 10.1038/ncomms15961} {\bibfield  {journal}
  {\bibinfo  {journal} {Nature Communications}\ }\textbf {\bibinfo {volume}
  {8}},\ \bibinfo {eid} {15961} (\bibinfo {year} {2017})}\BibitemShut {NoStop}%
\bibitem [{\citenamefont {{Kempkes}}\ \emph {et~al.}(2018)\citenamefont
  {{Kempkes}}, \citenamefont {{Slot}}, \citenamefont {{Freeney}}, \citenamefont
  {{Zevenhuizen}}, \citenamefont {{Vanmaekelbergh}}, \citenamefont {{Swart}},\
  and\ \citenamefont {{Smith}}}]{Kempkes}%
  \BibitemOpen
  \bibfield  {author} {\bibinfo {author} {\bibfnamefont {S.~N.}\ \bibnamefont
  {{Kempkes}}}, \bibinfo {author} {\bibfnamefont {M.~R.}\ \bibnamefont
  {{Slot}}}, \bibinfo {author} {\bibfnamefont {S.~E.}\ \bibnamefont
  {{Freeney}}}, \bibinfo {author} {\bibfnamefont {S.~J.~M.}\ \bibnamefont
  {{Zevenhuizen}}}, \bibinfo {author} {\bibfnamefont {D.}~\bibnamefont
  {{Vanmaekelbergh}}}, \bibinfo {author} {\bibfnamefont {I.}~\bibnamefont
  {{Swart}}}, \ and\ \bibinfo {author} {\bibfnamefont {C.~M.}\ \bibnamefont
  {{Smith}}},\ }\href@noop {} {\bibfield  {journal} {\bibinfo  {journal}
  {Nature Physics}\ }\textbf {\bibinfo {volume}
  {https://doi.org/10.1038/s41567-018-0328-0}} (\bibinfo {year}
  {2018})}\BibitemShut {NoStop}%
\bibitem [{\citenamefont {{Nurul Huda}}\ \emph {et~al.}(2018)\citenamefont
  {{Nurul Huda}}, \citenamefont {{Kezilebieke}}, \citenamefont {{Ojanen}},
  \citenamefont {{Drost}},\ and\ \citenamefont {{Liljeroth}}}]{Huda}%
  \BibitemOpen
  \bibfield  {author} {\bibinfo {author} {\bibfnamefont {M.}~\bibnamefont
  {{Nurul Huda}}}, \bibinfo {author} {\bibfnamefont {S.}~\bibnamefont
  {{Kezilebieke}}}, \bibinfo {author} {\bibfnamefont {T.}~\bibnamefont
  {{Ojanen}}}, \bibinfo {author} {\bibfnamefont {R.}~\bibnamefont {{Drost}}}, \
  and\ \bibinfo {author} {\bibfnamefont {P.}~\bibnamefont {{Liljeroth}}},\
  }\href@noop {} {\bibfield  {journal} {\bibinfo  {journal} {ArXiv e-prints}\ }
  (\bibinfo {year} {2018})},\ \Eprint {http://arxiv.org/abs/1806.08614}
  {arXiv:1806.08614 [cond-mat.mes-hall]} \BibitemShut {NoStop}%
\bibitem [{\citenamefont {{Ma}}\ \emph {et~al.}(2017)\citenamefont {{Ma}},
  \citenamefont {{Qiu}}, \citenamefont {{L{\"u}}},\ and\ \citenamefont
  {{Gao}}}]{Ma}%
  \BibitemOpen
  \bibfield  {author} {\bibinfo {author} {\bibfnamefont {L.}~\bibnamefont
  {{Ma}}}, \bibinfo {author} {\bibfnamefont {W.-X.}\ \bibnamefont {{Qiu}}},
  \bibinfo {author} {\bibfnamefont {J.-T.}\ \bibnamefont {{L{\"u}}}}, \ and\
  \bibinfo {author} {\bibfnamefont {J.-H.}\ \bibnamefont {{Gao}}},\ }\href@noop
  {} {\bibfield  {journal} {\bibinfo  {journal} {ArXiv e-prints}\ } (\bibinfo
  {year} {2017})},\ \Eprint {http://arxiv.org/abs/1707.04756} {arXiv:1707.04756
  [cond-mat.mes-hall]} \BibitemShut {NoStop}%
\bibitem [{\citenamefont {Paavilainen}\ \emph {et~al.}(2016)\citenamefont
  {Paavilainen}, \citenamefont {Ropo}, \citenamefont {Nieminen}, \citenamefont
  {Akola},\ and\ \citenamefont {Räsänen}}]{Paavilainen}%
  \BibitemOpen
  \bibfield  {author} {\bibinfo {author} {\bibfnamefont {S.}~\bibnamefont
  {Paavilainen}}, \bibinfo {author} {\bibfnamefont {M.}~\bibnamefont {Ropo}},
  \bibinfo {author} {\bibfnamefont {J.}~\bibnamefont {Nieminen}}, \bibinfo
  {author} {\bibfnamefont {J.}~\bibnamefont {Akola}}, \ and\ \bibinfo {author}
  {\bibfnamefont {E.}~\bibnamefont {Räsänen}},\ }\href@noop {} {\bibfield
  {journal} {\bibinfo  {journal} {Nano Letters}\ }\textbf {\bibinfo {volume}
  {16}},\ \bibinfo {pages} {3519} (\bibinfo {year} {2016})}\BibitemShut
  {NoStop}%
\bibitem [{\citenamefont {Stroscio}\ and\ \citenamefont
  {Eigler}(1991)}]{Stroscio1991}%
  \BibitemOpen
  \bibfield  {author} {\bibinfo {author} {\bibfnamefont {J.~A.}\ \bibnamefont
  {Stroscio}}\ and\ \bibinfo {author} {\bibfnamefont {D.~M.}\ \bibnamefont
  {Eigler}},\ }\href {\doibase 10.1126/science.254.5036.1319} {\bibfield
  {journal} {\bibinfo  {journal} {Science}\ }\textbf {\bibinfo {volume}
  {254}},\ \bibinfo {pages} {1319} (\bibinfo {year} {1991})}\BibitemShut
  {NoStop}%
\bibitem [{\citenamefont {Bartels}\ \emph {et~al.}(1997)\citenamefont
  {Bartels}, \citenamefont {Meyer},\ and\ \citenamefont {Rieder}}]{Bartels}%
  \BibitemOpen
  \bibfield  {author} {\bibinfo {author} {\bibfnamefont {L.}~\bibnamefont
  {Bartels}}, \bibinfo {author} {\bibfnamefont {G.}~\bibnamefont {Meyer}}, \
  and\ \bibinfo {author} {\bibfnamefont {K.-H.}\ \bibnamefont {Rieder}},\
  }\href@noop {} {\bibfield  {journal} {\bibinfo  {journal} {Phys. Rev. Lett.}\
  }\textbf {\bibinfo {volume} {79}},\ \bibinfo {pages} {697} (\bibinfo {year}
  {1997})}\BibitemShut {NoStop}%
\bibitem [{\citenamefont {Celotta}\ \emph {et~al.}(2014)\citenamefont
  {Celotta}, \citenamefont {Balakirsky}, \citenamefont {Fein}, \citenamefont
  {Hess}, \citenamefont {Rutter},\ and\ \citenamefont {Stroscio}}]{Celotta}%
  \BibitemOpen
  \bibfield  {author} {\bibinfo {author} {\bibfnamefont {R.~J.}\ \bibnamefont
  {Celotta}}, \bibinfo {author} {\bibfnamefont {S.~B.}\ \bibnamefont
  {Balakirsky}}, \bibinfo {author} {\bibfnamefont {A.~P.}\ \bibnamefont
  {Fein}}, \bibinfo {author} {\bibfnamefont {F.~M.}\ \bibnamefont {Hess}},
  \bibinfo {author} {\bibfnamefont {G.~M.}\ \bibnamefont {Rutter}}, \ and\
  \bibinfo {author} {\bibfnamefont {J.~A.}\ \bibnamefont {Stroscio}},\
  }\href@noop {} {\bibfield  {journal} {\bibinfo  {journal} {Review of
  Scientific Instruments}\ }\textbf {\bibinfo {volume} {85}},\ \bibinfo {pages}
  {121301} (\bibinfo {year} {2014})}\BibitemShut {NoStop}%
\bibitem [{\citenamefont {Park}\ and\ \citenamefont {Louie}(2009)}]{Park}%
  \BibitemOpen
  \bibfield  {author} {\bibinfo {author} {\bibfnamefont {C.-H.}\ \bibnamefont
  {Park}}\ and\ \bibinfo {author} {\bibfnamefont {S.~G.}\ \bibnamefont
  {Louie}},\ }\href@noop {} {\bibfield  {journal} {\bibinfo  {journal} {Nano
  Letters}\ }\textbf {\bibinfo {volume} {9}},\ \bibinfo {pages} {1793}
  (\bibinfo {year} {2009})}\BibitemShut {NoStop}%
\bibitem [{Sup()}]{Supp}%
  \BibitemOpen
  \href@noop {} {\bibinfo  {journal} {See Supplemental Material at \url{https://journals.aps.org/prx/supplemental/10.1103/PhysRevX.9.011009}}\
  }\BibitemShut {NoStop}%
\bibitem [{\citenamefont {Weeks}\ and\ \citenamefont {Franz}(2010)}]{Weeks}%
  \BibitemOpen
\bibfield  {journal} {  }\bibfield  {author} {\bibinfo {author} {\bibfnamefont
  {C.}~\bibnamefont {Weeks}}\ and\ \bibinfo {author} {\bibfnamefont
  {M.}~\bibnamefont {Franz}},\ }\href {\doibase 10.1103/PhysRevB.82.085310}
  {\bibfield  {journal} {\bibinfo  {journal} {Phys. Rev. B}\ }\textbf {\bibinfo
  {volume} {82}},\ \bibinfo {pages} {085310} (\bibinfo {year}
  {2010})}\BibitemShut {NoStop}%
\bibitem [{\citenamefont {Qiu}\ \emph {et~al.}(2016)\citenamefont {Qiu},
  \citenamefont {Li}, \citenamefont {Gao}, \citenamefont {Zhou},\ and\
  \citenamefont {Zhang}}]{Qiu}%
  \BibitemOpen
  \bibfield  {author} {\bibinfo {author} {\bibfnamefont {W.-X.}\ \bibnamefont
  {Qiu}}, \bibinfo {author} {\bibfnamefont {S.}~\bibnamefont {Li}}, \bibinfo
  {author} {\bibfnamefont {J.-H.}\ \bibnamefont {Gao}}, \bibinfo {author}
  {\bibfnamefont {Y.}~\bibnamefont {Zhou}}, \ and\ \bibinfo {author}
  {\bibfnamefont {F.-C.}\ \bibnamefont {Zhang}},\ }\href {\doibase
  10.1103/PhysRevB.94.241409} {\bibfield  {journal} {\bibinfo  {journal} {Phys.
  Rev. B}\ }\textbf {\bibinfo {volume} {94}},\ \bibinfo {pages} {241409}
  (\bibinfo {year} {2016})}\BibitemShut {NoStop}%
\bibitem [{\citenamefont {Menezes}\ \emph {et~al.}(2018)\citenamefont
  {Menezes}, \citenamefont {Smith},\ and\ \citenamefont {Palumbo}}]{Menezes}%
  \BibitemOpen
  \bibfield  {author} {\bibinfo {author} {\bibfnamefont {N.}~\bibnamefont
  {Menezes}}, \bibinfo {author} {\bibfnamefont {C.~M.}\ \bibnamefont {Smith}},
  \ and\ \bibinfo {author} {\bibfnamefont {G.}~\bibnamefont {Palumbo}},\ }\href
  {\doibase 10.1103/PhysRevB.97.075135} {\bibfield  {journal} {\bibinfo
  {journal} {Phys. Rev. B}\ }\textbf {\bibinfo {volume} {97}},\ \bibinfo
  {pages} {075135} (\bibinfo {year} {2018})}\BibitemShut {NoStop}%
\bibitem [{\citenamefont {Saito}\ \emph {et~al.}(1998)\citenamefont {Saito},
  \citenamefont {Dresselhaus},\ and\ \citenamefont
  {Dresselhaus}}]{Dresselhaus}%
  \BibitemOpen
  \bibfield  {author} {\bibinfo {author} {\bibfnamefont {R.}~\bibnamefont
  {Saito}}, \bibinfo {author} {\bibfnamefont {G.}~\bibnamefont {Dresselhaus}},
  \ and\ \bibinfo {author} {\bibfnamefont {M.}~\bibnamefont {Dresselhaus}},\
  }\href@noop {} {\emph {\bibinfo {title} {{Physical Properties of Carbon
  Nanotubes}}}}\ (\bibinfo  {publisher} {Imperial College Press},\ \bibinfo
  {year} {1998})\BibitemShut {NoStop}%
\bibitem [{\citenamefont {Tadjine}\ \emph {et~al.}(2016)\citenamefont
  {Tadjine}, \citenamefont {Allan},\ and\ \citenamefont {Delerue}}]{Tadjine}%
  \BibitemOpen
  \bibfield  {author} {\bibinfo {author} {\bibfnamefont {A.}~\bibnamefont
  {Tadjine}}, \bibinfo {author} {\bibfnamefont {G.}~\bibnamefont {Allan}}, \
  and\ \bibinfo {author} {\bibfnamefont {C.}~\bibnamefont {Delerue}},\ }\href
  {\doibase 10.1103/PhysRevB.94.075441} {\bibfield  {journal} {\bibinfo
  {journal} {Phys. Rev. B}\ }\textbf {\bibinfo {volume} {94}},\ \bibinfo
  {pages} {075441} (\bibinfo {year} {2016})}\BibitemShut {NoStop}%
\bibitem [{\citenamefont {Crommie}\ \emph {et~al.}(1993)\citenamefont
  {Crommie}, \citenamefont {Lutz},\ and\ \citenamefont {Eigler}}]{Crommie}%
  \BibitemOpen
  \bibfield  {author} {\bibinfo {author} {\bibfnamefont {M.~F.}\ \bibnamefont
  {Crommie}}, \bibinfo {author} {\bibfnamefont {C.~P.}\ \bibnamefont {Lutz}}, \
  and\ \bibinfo {author} {\bibfnamefont {D.~M.}\ \bibnamefont {Eigler}},\
  }\href@noop {} {\bibfield  {journal} {\bibinfo  {journal} {Science}\ }\textbf
  {\bibinfo {volume} {262}},\ \bibinfo {pages} {218} (\bibinfo {year}
  {1993})}\BibitemShut {NoStop}%
\end{thebibliography}
\end{document}